\documentclass[fleqn,usenatbib]{mnras}

\usepackage{newtxtext,newtxmath}

\usepackage[T1]{fontenc}

\DeclareRobustCommand{\VAN}[3]{#2}
\let\VANthebibliography\thebibliography
\def\thebibliography{\DeclareRobustCommand{\VAN}[3]{##3}\VANthebibliography}

\usepackage{graphicx}	
\usepackage{amsmath}	
\usepackage{tabularx}
\defcitealias{Weltevrede2006}{W06}
\defcitealias{Malofeev2018}{M18}

\title{SEARCH FOR WEAK COMPONENTS IN PULSAR RADIATION}

\author[T. V. Smirnova et al.]{
T. V. Smirnova $^{a}$\thanks{E-mail: tania@prao.ru}
M. O. Toropov,$^{b}$
S. A. Tyul'bashev,$^{a}$
\\
$^{a}$ Lebedev Physical Institute, Astro Space Center, Pushchino Radio Astronomy Observatory, Russian Academy of Sciences, Pushchino, 142290 Russia \\
$^{b}$ LLC TEK Inform, Moscow, 117246 Russia \\
}

\date{2024}

\pubyear{2024}

\begin{document}
\label{firstpage}
\pagerange{\pageref{firstpage}--\pageref{lastpage}}
\maketitle

\begin{abstract}
The search for weak components outside the main pulse (MP) in the radiation of pulsars at a frequency of 110 MHz observed on the LPA LPI telescope in the Pushchino Multibeam Pulsar Search (PUMPS) has been carried out. The sample included 96 pulsars, for which the signal-to-noise ratio (S/N) in the MP of the average profile during accumulation over 10 years was more than 40. It was found that PSR J1543+0929 has radiation for almost the entire period. The profile is three-component. The relative amplitudes of the lateral weak components are 0.013 and 0.026. For PSR J2234+2114, a precursor was detected that is $53^{\circ}$ away from MP.
\end{abstract}




\section{Introduction}
Most pulsars have radio emission in a narrow cone and the typical width of the average profiles obtained by the accumulation of several hundred or even thousands of pulses is 5-10\% of the pulsar period ($P$). Although individual pulses have different shapes, the average profile for a given pulsar is usually very stable at a given frequency  (\citeauthor{Helfand1975}, \citeyear{Helfand1975}),  (\citeauthor{Rankin1995}, \citeyear{Rankin1995}). However, there are pulsars with mode switching (PSR B0943+10, B1237+25 and others) when the profile shape changes discretely  (\citeauthor{Taylor1975}, \citeyear{Taylor1975}). The observed shape of the pulse depends significantly on the size and structure of the emitting region, as well as on the angle between the direction towards the observer and the center of the radiation cone.

The observed shapes of the average profiles can be either single-component (for example, for PSR B1933+16) or multicomponent (PSR B1237+25 has 5 components). In very rare cases, pulsars show radiation throughout the entire period. Thus, the pulsar B0826-34 has radiation in the region of 250 degrees (69\% of the period) at meter waves and for at least 70\% of the time its radiation is not visible (nullings)  (\citeauthor{Durdin1979}, \citeyear{Durdin1979}),  (\citeauthor{Biggs1985}, \citeyear{Biggs1985}). For the pulsar B0950+08, radiation is also detected for almost the entire period  (\citeauthor{Hankins1981}, \citeyear{Hankins1981}),  (\citeauthor{Smirnova1988}, \citeyear{Smirnova1988}),  (\citeauthor{Smirnova2024}, \citeyear{Smirnova2024}). This assumes that the angle between the magnetic axis and the axis of rotation is small (coaxial rotator) and the direction towards the observer is always inside the radiation region. A number of pulsars have pulses at a distance of about 180$^{\circ}$ from the main pulse (MP). These pulses are called interpulses (IP) and they usually have a significantly lower amplitude compared to MP. Pulsars with IP can be both coaxial and orthogonal rotators. In the case of an orthogonal rotator, the observer can see the radiation from both magnetic poles. In order to distinguish the case of coaxial or orthogonal rotators, polarization measurements can be used (\citeauthor{Olszanski2019}, \citeyear{Olszanski2019}).

The purpose of this work is to search for weak components in the radio emission of pulsars outside the main pulse from a sample of pulsars observed in the Pushchino Multibeam Pulsar Search (PUMPS)  (\citeauthor{Tyul'bashev2022}, \citeyear{Tyul'bashev2022}).

\begin{figure*}
\begin{center}
	\includegraphics[width=0.7\textwidth]{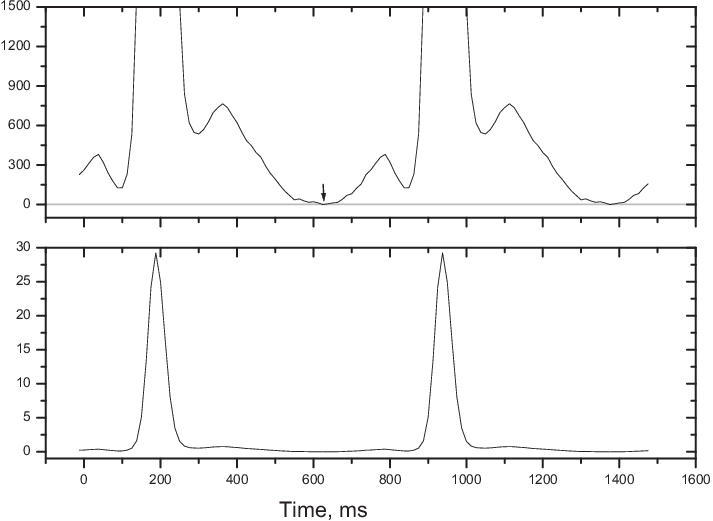}
    \caption{The average profiles accumulated over 10 years (2883 sessions) of the pulsar J1543+0929 are given with a double period. At the bottom is the original profile, at the top is the same profile with an amplitude increased by 1000 times. The precursor and postcursor are clearly visible in the upper part of the figure. The arrow shows the minimum radiation occupying several points in the profile. The x-axis is the time in ms, the y-axis is the amplitude of the profile in relative units.}
    \label{fig:fig1}
\end{center}
\end{figure*}

\section{OBSERVATIONS AND DATA PROCESSING}

Monitoring observations are carried out on the Large Phased Array (LPA3) of the Lebedev Physical Institute (LPI) under the pulsar and transient search program starting in August 2014. We used this data in our work. The antenna receives linearly polarized radiation. The central frequency of observations is 110.3 MHz, the bandwidth is 2.5 MHz. 128 antenna beams are aligned in the plane of the meridian and cover declinations from $-9^{\circ}$ to $+55^{\circ}$ with an overlap of beams at the level of 0.4. The time of passage through the meridian at half power (observation session) is about 3.5 min. In this work, a 32-channel receiver with a channel width of 78 kHz was used. The sampling of the point is 12.5 ms. To calibrate the signal in the frequency channels, a calibration step with a known temperature (the signal from the noise generator) was used, which was recorded 6 times a day. According to the calibration steps, during further processing, the signal gain was equalized in different sessions and in all frequency channels  (\citeauthor{Tyul'bashev2019}, \citeyear{Tyul'bashev2019}). Over 10 years of round-the-clock monitoring, about 3,000 observation sessions have been accumulated for each pulsar.

The data was recorded to the hard disk in hourly portions in all frequency channels. The part corresponding to the time of passage of this pulsar through the antenna radiation pattern at the half-power level was selected from the hourly recording. The primary data processing included: calibration of the signal according to the calibration step so that the gain in all channels was the same, subtraction of the baseline, compensation of the dispersion measure ($DM$), recording of all pulses in all channels on a disk.

The average profile for each session was obtained after compensation of $DM$ by adding all pulse records with a given period of $P$. If the signal-to-noise ratio ($S/N$) in the main pulse was less than 8, the session was discarded. For the remaining sessions, a cyclical shift was carried out in each average profile so that the main pulse was in the first quarter of the period. Thus, the MP phase was the same in all sessions, so it was possible to add up the average profiles for each session to get the average profile for all selected sessions of a given year. An estimate of the standard deviation (rms) noise outside the pulse ($\sigma_n$) was made. In addition, files were generated with recordings of the pulses of each session and the average profile for this session. Then the profiles of all sessions for each year were summed up and we got the average profiles by year, as well as the summarized average profile for all years. To analyze the data, a sample of 96 pulsars was used, for which the ratio of $S/N$ in the main pulse of the average profile during accumulation over 10 years (2014-2023) was more than 40. Table 1 of  (\citeauthor{Toropov2024}, \citeyear{Toropov2024}) shows the names of these pulsars, $P$ and $DM$.

\section{RESULTS AND DISCUSSION}

Analysis of the profiles of 96 pulsars obtained over 10 years of observations made it possible to detect 12 pulsars with IP, and in 9 pulsars IP were detected for the first time  (\citeauthor{Toropov2024}, \citeyear{Toropov2024}). No new details (features) were found in the summarized average profiles of 82 pulsars. Two pulsars (J1543+0929 and J2234+2114) have unusual details in their profiles, and we analyze them in this work.

\begin{figure*}
\begin{center}
	\includegraphics[width=0.7\textwidth]{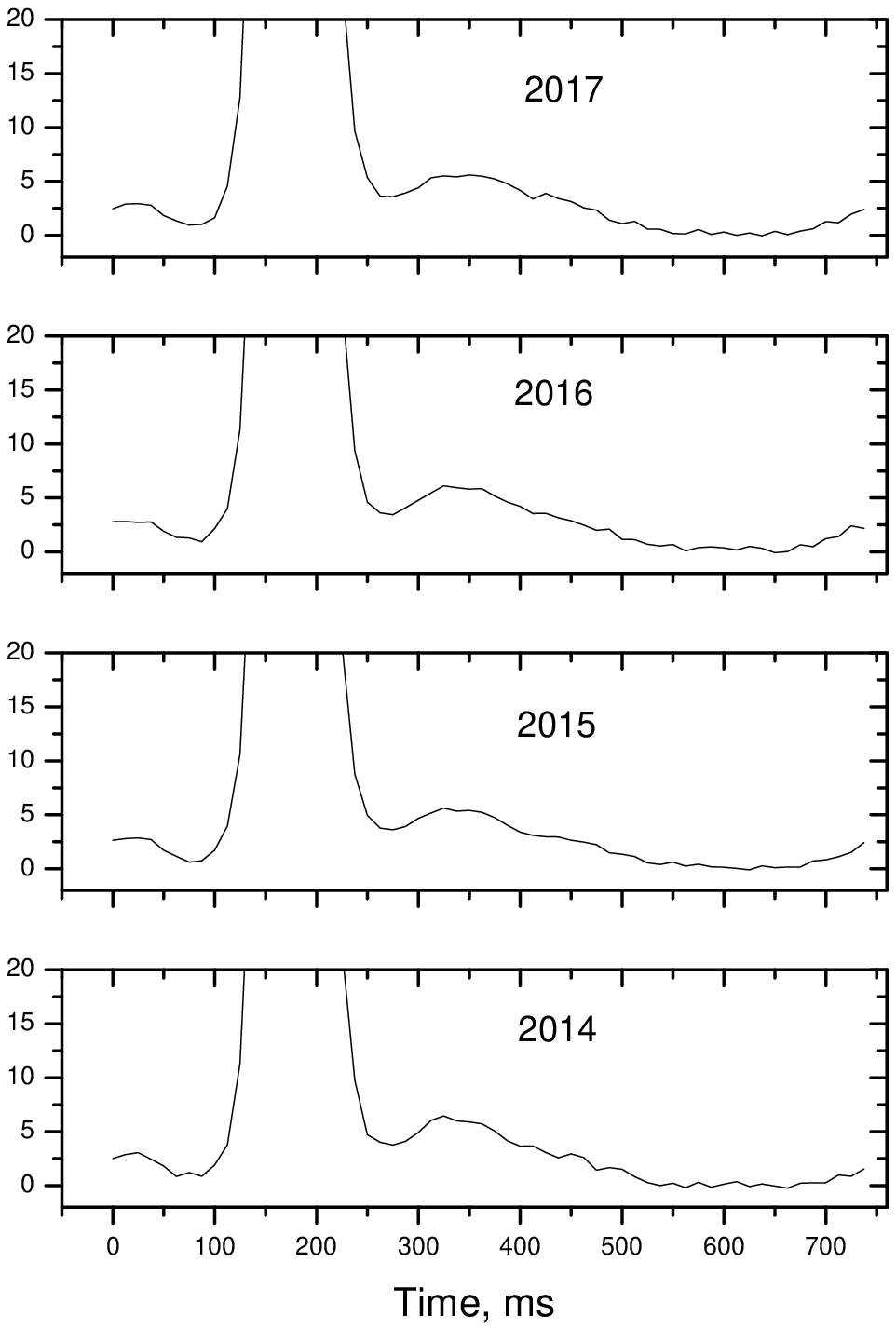}
    \caption{The average profiles accumulated over each of the four years are J1543+0929. The x and y axes correspond to the designations in Fig. 1.}
    \label{fig:fig2}
\end{center}
\end{figure*}

\textit{J1543+0929}. Pulsar J1543+0929 (B1541+09) has radiation detected almost throughout the entire period. Fig.~\ref{fig:fig1} shows the summarized average profile of J1543+0929 for all years of observation with a double period. It can be seen that only in a small part of the average profile (less than 10\% of the total period) there is no radiation. The minimum marked with an arrow in the figure is located at a distance of $115^{\circ}$ (638 ms) from MP. The line of the average profile looks very smooth due to the averaging of a large number of profiles. The value of the standard deviation on the summed profile is comparable to the width of the line. The resulting pattern of radiation observed in almost the entire period of J1543+0929 is not accidental. Fig.~\ref{fig:fig2} shows the average profiles for four consecutive years. All profiles accumulated over the year are stable and have the same minimum amplitude region with a duration of approximately 70 ms. For 2018-2023, the profiles have a similar appearance. There is no temporal evolution of the average profile from year to year. This pulsar also has two side components: a pre-pulse (precursor) and a post-pulse (post-cursor).

\begin{figure*}
\begin{center}
	\includegraphics[width=0.6\textwidth]{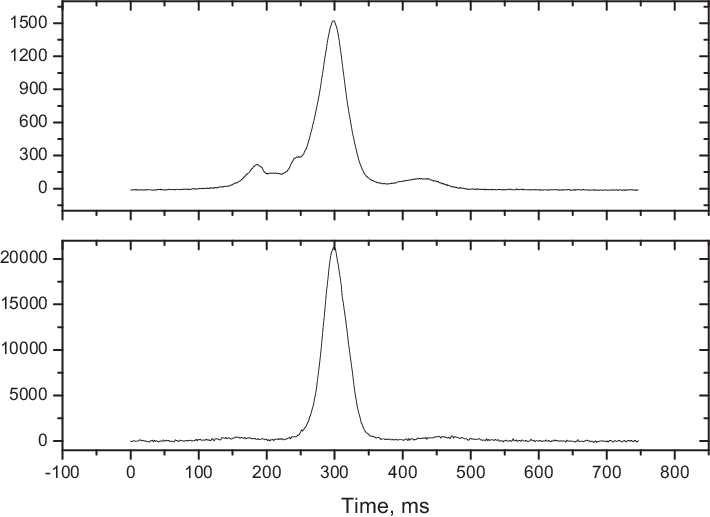}
    \caption{The average profiles of J1543+0929 at frequencies of 139.2 MHz (bottom) and 327 MHz (top), taken from the pulsar profile database https://psrweb.jb.man.ac.uk/epndb . The x-axis is the time in ms, the y-axis is the amplitude in relative units.}
    \label{fig:fig3}
\end{center}
\end{figure*}

Previously, the average profiles of J1543+0929 were obtained by different authors in a wide frequency range: from 65 MHz to 4750 MHz and are shown in the pulsar profile database (https://psrweb.jb.man.ac.uk/epndb ). For comparison, we present profiles at a frequency of 139.2 MHz (\citeauthor{Bilous2016}, \citeyear{Bilous2016}) and at a frequency of 327 MHz (\citeauthor{Olszanski2019}, \citeyear{Olszanski2019}) taken from this database. They are shown in Fig.~\ref{fig:fig3}. At 139.2 MHz, $t_1 = -140$~ms ($-67.4^{\circ}$), $t_2 = +155$~ms ($+74.6^{\circ}$), $A_1 = 0.02$, $A_2 = 0.025$  (\citeauthor{Bilous2016}, \citeyear{Bilous2016}). At this frequency, the distances from the MP: $t_1$ and $t_2$ coincide with the one we obtained with accuracy to measurement errors, given that our resolution is 12.5 ms and the dispersion smearing in the channel band is 17.5 ms. The ratio of the amplitudes of the components is approximately the same. At a frequency of 327 MHz, the relative amplitudes increase and $A_2$/$A_1 = 2.3$, and the distances from the MP decrease: $t_1 = -111$~ms ($-53.4^{\circ}$), $t_2 = +134$~ms ($+64.5^{\circ}$)  (\citeauthor{Olszanski2019}, \citeyear{Olszanski2019}). An increase in the distance between the MP and the side components with a decrease in frequency is usually associated with the fact that radiation at lower frequencies comes from higher levels from the surface of a neutron star, which corresponds to an expansion of the cone of magnetic field lines  (\citeauthor{Rankin1983}, \citeyear{Rankin1983}) and, consequently, an increase in the distance between the components.

At low frequencies, the J1543+0929 profile is 3-component and corresponds to the "core-cone" model of the Rankin classification (\citeauthor{Rankin1993}, \citeyear{Rankin1993}), which considers the cone-shaped structure of the radiating region, in the center of which the central region ("core") is located. The work (\citeauthor{Olszanski2019}, \citeyear{Olszanski2019}) investigated the polarization properties of 46 pulsars. This list also included the pulsar J1543+0929. The analysis of the widths of the profile components and the behavior of the polarization angle with longitude allowed in this work to estimate the angle between the magnetic axis and the axis of rotation $\alpha$ and the impact angle $\beta$, under which the line on the observer cuts the radiation cone. For J1543+0929, the angle is $\alpha$ = $6^\circ$ and $\beta$ = $-2.20^\circ$. The angle $\beta$ is small and this means almost the central section of the radiation cone, the observer sees all three components of the profile. Since the angle between the magnetic axis and the axis of rotation is small, the observer can see the radiation for almost the entire period. However, this radiation for J1543+0929 was first detected only in this work.

The discovery of IP radiation is primarily due to the very high $S/N$ in the average profile accumulated over many years. The level of the detected pulse-to-pulse radiation signal is several times lower than the signal level in the precursor and is a fraction of a percent of the MP signal. The same effect is observed for the pulsar B0953+09, which was noted in the Introduction.

\begin{figure*}
\begin{center}
	\includegraphics[width=0.6\textwidth, angle=270]{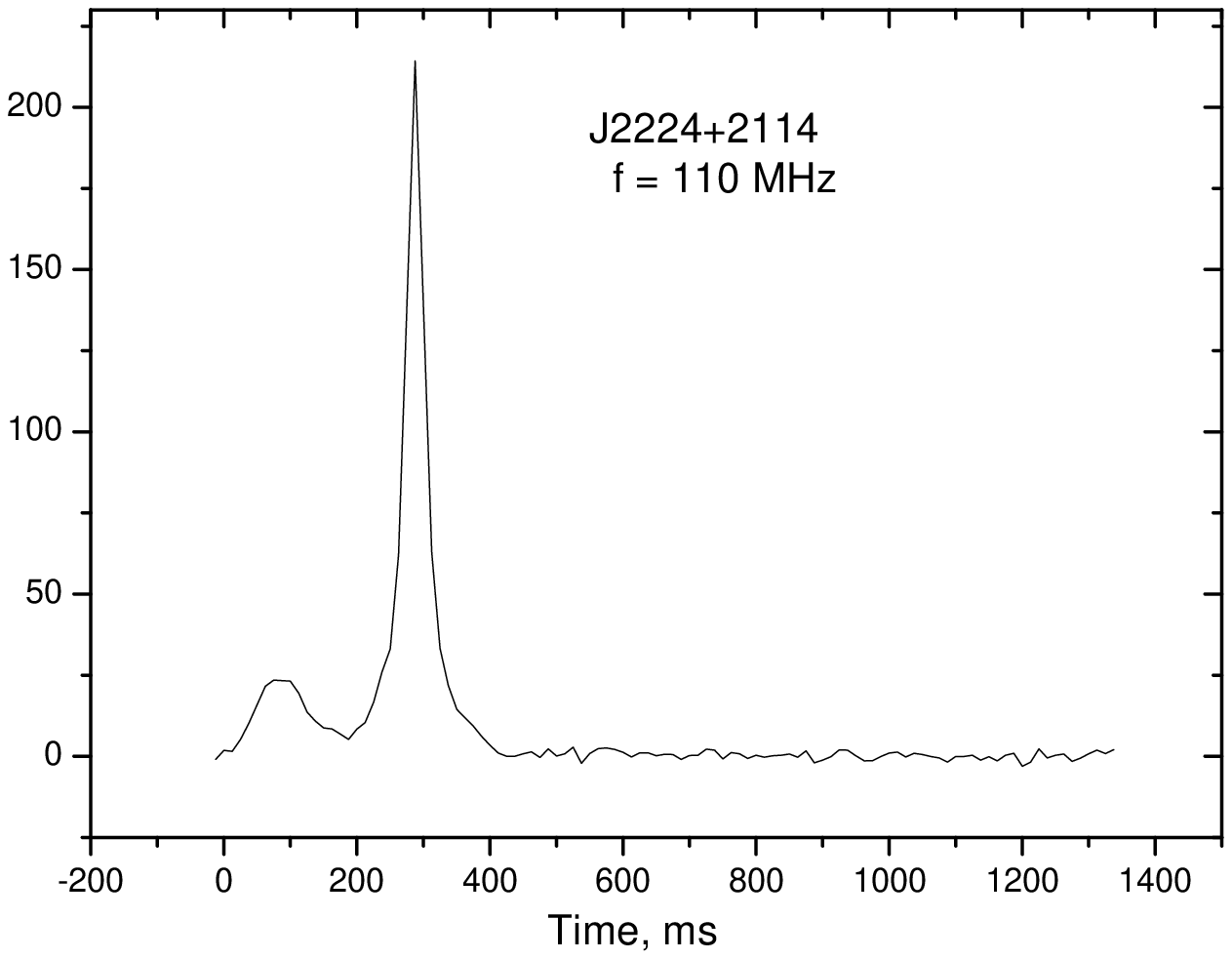}
    \caption{Average profiles accumulated over 10 years (3101 sessions) J2234+2114. The x-axis is the time in ms, the y-axis is the amplitude of the profile in relative units.}
    \label{fig:fig4}
\end{center}
\end{figure*}

\textit{J2234+2114.} A precursor has been detected in pulsar J2234+2114, which was not previously described. The total average profile over 10 years of observations for the pulsar J2234+2114 is shown in Fig.~\ref{fig:fig4}. It has a pre-pulse located at a distance of -200 ms ($-53^{\circ}$) from the MP and a relative amplitude of 0.1.

There are only 3 profiles for this pulsar in the pulsar profile database (https://psrweb.jb.man.ac.uk/epndb ): at frequencies of 148.9 MHz (\citeauthor{Bilous2016}, \citeyear{Bilous2016}), 346.7 MHz (\citeauthor{Wahl2023}, \citeyear{Wahl2023}) and 430 MHz  (\citeauthor{Camilo1995}, \citeyear{Camilo1995}). At a frequency of 148.9 MHz, the profile is single-component with a low $S/N$ ratio, and at 346.7 MHz and 430 MHz, the profiles are apparently 3-component with components passing into one another without a clear separation. The distance between the leftmost component and the maximum of the profile is difficult to determine, but the ratio of its amplitude to the amplitude of the maximum of the profile coincides with the value at 110 MHz. Perhaps this component corresponds to our pre-pulse.

Table~\ref{tab:tab1} shows the parameters of the studied pulsars: $P$, $DM$, the number of sessions involved in the accumulation of $N$ over 10 years of observations, the distance of the components before (the "-" sign) and after (the "+" sign) of MP maximum in ms (in these distances are given in parentheses in degrees), $t_1$ and $t_2$, respectively, as well as their relative amplitudes $A_1$ and $A_2$. The accuracy of determining the distance of the components from the MP in ms and degrees corresponds to approximately one point of the average profile ($\pm 12.5$~ms) or $\pm 6^{\circ}$.

\begin{table*}
\centering
\caption{The amplitude ratios of the components of the B0950+08 profile and their peak flow densities by year.}
\label{tab:tab1}
\begin{tabular}{cccccccc}
\hline
PSR &  $P$, ms  &  $DM$, pc/cm$^3$ & N & $t_1$, ms & $t_2$, ms & $A_1$ & $A_2$ \\
\hline
J1543+0929 & 0.748 & 34.976 & 2883 & -150 ($-72^{\circ}$) & +175 ($+84^{\circ}$) & 0.013 & 0.026\\
J2234+2114 & 1.359 & 35.08 & 3101 & -200 ($-53^{\circ}$) & - & - & 0.11\\
\hline
\end{tabular}
\label{tab:tab1}
\end{table*}

As can be seen from column 4, indicating the number of sessions for which average profiles were obtained, the growth of $S/N$ in the main pulse when all profiles are added together will be approximately 55 times greater than in one session. Considering that the effective area of the LPA in the meter range is the highest in the world, the additional growth of $S/N$ leads to the identification of new details in the average profiles of long-known and well-studied pulsars.

\section{CONCLUSIONS}

As a result of the analysis of the summarized average profiles of pulsars J1543+0929 and J2234+2114 over 10 years, it was possible to detect features that had not been previously noted in the literature.

The pulsar PSR J1543+0929 has low-level radiation (inter-pulse radiation) at almost all longitudes within the pulsar period. Apparently, this is only the second pulsar (after PSR J0953+0755), the radiation of which takes more than 90\% of the period. Most likely, this radiation is due to the fact that the pulsar is coaxial. Its angle between the axis of rotation and the magnetic axis is small and therefore the observer can see the radiation area during the entire period.

A pre-pulse was detected in PSR J2234+2114, which is separated from MP at a distance of $53^{\circ}$ and has a relative amplitude of 0.1. At frequencies 346.7 MHz and 430 MHz, the profile has a more complex appearance, apparently with three components.

\section*{ACKNOWLEDGMENTS}

The authors thank the LPA antenna group for the constant support of the radio telescope in monitoring mode, and L.B. Potapova for help in preparing the paper.

\bsp	
\label{lastpage}

\bibliographystyle{mnras}
\bibliography{sts}

\end{document}